\documentclass[authoryear,preprint,review,12pt]{elsarticle}

\usepackage{graphicx}

\usepackage{amssymb}

\journal{J. Atm. Sol.-Terr. Phys.}

\begin{document}

\begin{frontmatter}



\title{Towards a long-term record of solar total and spectral irradiance}


\author[i1]{N.\,A.~Krivova}
\author[i1,i2]{S.\,K.~Solanki}
\author[i3]{Y.\,C.~Unruh}
\address[i1]{Max-Planck-Institut f\"ur Sonnensystemforschung,
Max-Planck-Str. 2, 37191 Katlenburg-Lindau, Germany}
\address[i2]{School of Space Research, Kyung Hee University,
Yongin, Gyeonggi 446-701, Korea}
\address[i3]{Astrophysics Group, Blackett Laboratory, Imperial College
London, SW7 2AZ, United Kingdom}

\begin{abstract}
The variation of total solar irradiance (TSI) has been measured since 1978
and that of the spectral irradiance for an even shorter amount of time.
Semi-empirical models are now available that
reproduce over 80\% of the measured irradiance variations.
An
extension of these models into the more distant past is needed in order to
serve as input to climate simulations.
Here we review our most recent efforts to model solar total and
spectral irradiance on time scales from days to centuries and even longer.
Solar spectral irradiance has been reconstructed since 1947.
Reconstruction of solar total irradiance goes back to 1610 and suggests a
value of about 1--1.5~Wm$^{-2}$ for the increase in the cycle-averaged TSI
since the end of the Maunder minimum, which is significantly lower than
previously assumed but agrees with other modern models.
First steps have also been made towards reconstructions of solar total and
spectral irradiance on time scales of millennia.
\end{abstract}


\end{frontmatter}


\section{Introduction}
\label{intro}

Solar irradiance is the total energy flux (or energy received per unit area
and time) at the top of the Earth's atmosphere.
It is strongly wavelength dependent: less than 8\% of the solar energy
is emitted at wavelengths below 400~nm, more than 60\% come from the
wavelengths between 400 and 1000~nm and roughly another 30\% from the longer
wavelengths \citep[e.g.,][]{solanki-unruh-98,krivova-et-al-2006a}.

The small contribution of the UV part of the spectrum is, however,
compensated by the spectral dependence of the transmission of the Earth's
atmosphere.
Thus solar radiation below 300~nm is almost completely absorbed in the
atmosphere \citep[see, e.g.,][and references therein]{haigh-2007} and is
important for the
chemistry of the stratosphere and overlying layers.
In particular, radiation in the Ly-$\alpha$ line (121.6~nm) and in the
oxygen continuum and bands between 180 and 240~nm controls production and
destruction of ozone
\citep[e.g.,][]{frederick-77,brasseur-simon-81,haigh-94,haigh-2007,%
fleming-et-al-95,egorova-et-al-2004,langematz-et-al-2005b}.
Solar UV radiation at 200--350~nm is the main heat source in the
stratosphere and mesosphere \citep{haigh-99,haigh-2007,rozanov-et-al-2006}.

The role of the solar UV radiation for the Earth's climate is boosted
further by the fact that variations of solar irradiance are also strongly
wavelength dependent.
Whereas the total (integrated over all wavelengths) solar irradiance varies
by only about 0.1\% over the solar cycle, the UV irradiance varies by 1 to 3
orders of magnitude more \citep[e.g.,][]{floyd-et-al-2003a}.

Solar near-IR radiation absorbed by water vapour and carbon dioxide
is an important source of heating in the lower atmosphere
\citep{haigh-2007}.
Solar variability in the IR is comparable to or lower than the TSI variations
and in the range between about 1500 and 2500~nm it is reversed with respect
to the solar cycle \citep{harder-et-al-2009,krivova-et-al-2009a}.

Solid assessment of the solar forcing on the Earth's climate is still
plagued, among other factors, by a shortage of reliable and sufficiently long
irradiance records.
The time series of direct space-borne measurements of solar irradiance is only
3 decades long (in case of spectral irradiance, even shorter).
Consequently, models are required to extend the irradiance time series to
earlier times.
In the last years, considerable advances have been made in modelling the
solar total irradiance.
Also, spectral models started gathering force.
Reconstructions of the past variations are quite accurate for the last
decades, when sufficiently high resolution measurements of the solar
photospheric magnetic flux (i.e. magnetograms) exist.
On longer time scales, the secular trend remains rather uncertain and until
very recently the reconstructions were limited to the telescope era, i.e. to
the last 4 centuries.

\citet{domingo-et-al-2009} have recently presented an overview of different
types of irradiance models on time scales of days up to the solar cycle.
Here we review the most recent progress in modelling solar total and
spectral irradiance on time scales from days to centuries and even longer
using the SATIRE set of models.
Before discussing the model we briefly describe the available observational
data (Sect.~\ref{obs}) and main sources of irradiance variations
(Sect.~\ref{sources}).
In Sect.~\ref{short} we concentrate on the period when direct irradiance
measurements are available, in Sects.~\ref{cent} and \ref{mill} we discuss
the reconstructions for the telescope era and the first steps towards
longer-term reconstructions, respectively.
Section~\ref{summary} summarises our current knowledge on irradiance
variations on time scales of interest for climate research.

\section{Solar Irradiance Variations}
\label{solirr}

\subsection{Observations}
\label{obs}

\subsubsection{Total Solar Irradiance}
\label{obs_tsi}

Measurements of the total solar irradiance (TSI) from space, having
sufficient accuracy to detect its variations, started with the
Hickey-Frieden (HF) radiometer \citep{hickey-et-al-80} aboard the NOAA/NASA
mission Nimbus-7.
It was followed by the Active Cavity Radiometer for
Irradiance Monitoring (ACRIM)~I on SMM \citep{willson-et-al-81}, 
the Earth Radiation Budget Experiment (ERBE) on ERBS
\citep{lee-et-al-1995}, ACRIM-II on UARS \citep{willson-94}, Variability of
Irradiance and Gravity Oscillations (VIRGO) on SoHO
\citep{froehlich-et-al-97}, ACRIM-III on ACRIMSAT \citep{willson-2001} and
the Total Solar Irradiance Monitor (TIM) on SORCE \citep{kopp-lawrence-2005}.

The measurements disclosed that the value previously called the solar
constant in fact varied at various time scales, from minutes to decades.
The typical magnitude of variations in the last 3 cycles was about 0.1\%,
although changes up to about 0.3\% were recorded on shorter time scales
(days), e.g. around the Halloween storm at the end of October 2003.

Since none of the instruments survived over the whole period since 1978 and
each of them suffers from its individual degradation, calibration or other
problems, a construction of a composite TSI record is quite a challenge.
Therefore, 3 different composites are now available: from the
Physikalisch-Meteorologisches Observatorium Davos
\citep[PMOD;][]{froehlich-2006}, the ACRIM team
\citep{willson-mordvinov-2003} and the
Institut Royal Meteorologique Belgique \citep[IRMB][]{dewitte-et-al-2004}.

They show certain differences, of which the most important regards the
presence or absence of an upward secular trend between the minima preceeding
cycles 22 (1986) and 23 (1996).
This long-standing argument traces its origin back to the unforseen gap
between the first two ACRIM experiments (July 1989 to October 1991), known
as the ACRIM gap.
Although 2 other experiments, HF and ERBE, continued operating
during these 2 years, they are claimed to be lower precision experiments
by \citet{scafetta-willson-2009}.
Moreover, the ERBE made measurements only approximately every second week,
whereas in September 1989 the output signal of the HF radiometer saturated
and the instrument was switched off for several days.
Once back in operation, it showed an abrupt increase in irradiance by about
0.4~W/m$^2$.
Somewhat later another increase of similar magnitude was reported
\citep{lee-et-al-1995,chapman-et-al-96}.
Depending on whether this change is taken into account (as in the PMOD
composite) or not (ACRIM), the composite time series shows either no change
in the TSI average level from 1986 to 1996 \citep{froehlich-2006} or a
significant increase \citep{willson-mordvinov-2003}.

Another uncertainty of the TSI record regards the absolute level of the
irradiance.
The uncertainty in the absolute values of the TSI is significantly higher
than that of the relative changes.
In particular, there remains an unexplained difference of about 4.5~W/m$^2$
between SORCE/TIM measurements and other currently operating instruments
\citep{kopp-et-al-2005b}.
A comparison of the TIM measurements to the NIST (National Institute of
Standards and Technology) reference detectors showed a rather good
agreement thus providing support to TIM's lower values.
The future Glory mission of NASA promises to throw light on this, since
the TSI measured with the Glory/TIM instrument will, for the first time,
be compared directly to the TSI measurements by the reference cryogenic
radiometer under the same vacuum conditions.
For studies of climate change, variations in the TSI are of greater
importance than the exact absolute value.

\subsubsection{Spectral Irradiance}
\label{obs_ssi}

The record of spectral irradiance measurements, accurate enough to assess
its variations, goes back to the launch of the Solar Mesospheric Explorer
(SME) in 1981.
Although data between 115 and 302~nm were obtained, their accuracy is not
sufficiently high due to degradation problems.
This led to the introduction of the Mg~II core-to-wing index, which, being a
ratio of the fluxes measured at close-by wavelengths
\citep{heath-schlesinger-86,viereck-puga-99}, suffers significantly
less from instrument degradation.
Mg~II observations were continued by a number of following missions and a
composite time series was produced by \citet{viereck-et-al-2004}.

The Upper Atmosphere Research Satellite (UARS) launched in 1991 had 2
instruments for measurements of solar UV irradiance between approximately
115 and 420~nm, the Solar Ultraviolet Spectral Irradiance Monitor
\citep[SUSIM;][]{brueckner-et-al-93} and the Solar Stellar Comparison
Experiment \citep[SOLSTICE;][]{rottman-et-al-93}.
UARS SUSIM and SOLSTICE measurements revealed the strong spectral dependence
of the solar irradiance variations \citep[see, e.g., the review
by][]{floyd-et-al-2003a}. 
However, the variability above approximately 300~nm remained rather
uncertain due to insufficient long-term stability of both instruments.

Since 1996 measurements have also being taken in three spectral channels
centered at 402, 500 and 862~nm with the VIRGO Sun PhotoMeter (SPM) on SoHO
\citep{froehlich-et-al-97}.
The detectors are, however, not stable enough to allow an independent
estimate of the comparatively weak longer-term trends at these wavelengths.

Data in a broader spectral range became available with the launch of
the Solar Radiation and Climate Experiment \citep[SORCE;][]{rottman-2005}
and the Environmental Satellite (ENVISAT) in 2003.
SORCE has two instruments for measuring solar spectral irradiance: SOLSTICE
and the Spectral Irradiance Monitor (SIM) operating at 115--425~nm and
200--2700~nm, respectively \citep{snow-et-al-2005,harder-et-al-2005a}.
The SCanning Imaging Absorption spectroMeter for Atmospheric CHartographY
(SCIAMACHY) on ENVISAT observes in the range 240--2380~nm
\citep{skupin-et-al-2005}.
The data from SIM and SCIAMACHY provided, for the first time, information
about solar variability in the IR part of the spectrum and about longer-term
(time scales of years) variations above 300~nm \citep{harder-et-al-2009,
pagaran-et-al-2009}.
These observations cover only the descending phase of solar cycle ~23.
The data over broad spectral sub-intervals are available, although there are
still some remaining problems in some others (in particular, in the IR).

In 2008, the SOLar SPECtrum (SOLSPEC) instrument on SOLAR/ISS started
observations between 165~nm and 3080~nm
\citep{thuillier-et-al-2009}.
The data are not yet publicly available.

Note that here we only discuss the spectral range above 115~nm.
Variations of the irradiance at yet shorter wavelengths is of more interest
for space weather studies and their measurements and modelling have recently
been reviewed by \citet{woods-2008}.

\subsection{Sources of irradiance variations}
\label{sources}

Just a view of the solar surface with a modern telescope (and to some
extent even without) suggests that brightness of the solar disc is not
homogeneous and changes in time.
Sunspots are the largest magnetic structures on the solar surface and have
been seen already in antiquity with the naked eye.
Being cooler than the surrounding photosphere, they appear darker.
Thus a passage of a spot across the visible solar disc causes a dip in solar
brightness \citep[see, e.g.,][]{hudson-et-al-82,fligge-et-al-2000a}.
The spots appear particularly dark when they are close to the disc centre
when their projected area is largest and the dark umbra is best visible.
The related irradiance changes occur on time scales of hours to about
10~days.

Since the amount of and area covered by sunspots increases significantly
from activity minimum to cycle maximum, the frequency and depths of
brightness dips related to spots also increase considerably.
Some of the dips reach up to 3--4~W/m$^2$ (or about 0.2--0.3\%).
On average, however, the Sun brightens from cycle minima to maxima.
This is because of the contribution of faculae and the network, which are
weaker magnetic elements appearing bright on the solar surface
\citep{zwaan-78,solanki-93}.
Faculae are formed in active regions and typically appear in the vicinity of
sunspots.
The weaker network elements are spread over the whole disc, forming a bright
network at the intersections of supergranules.
In the visible continuum, the contrast of faculae is significantly higher
close to the limb.
Thus when a facular region crosses the visible solar disc it typically
causes a weakly double-peaked brightening \citep[see,
e.g.,][]{fligge-et-al-2000a}.

\citet{wenzler-2005} has considered the net contributions of sunspots and
faculae to the TSI variation of cycle 23.
He found that sunspots caused a darkening of about 0.84~W/m$^2$, whereas
faculae produced a brightening of about 1.73~W/m$^2$.
The net increase in the TSI thus was 0.89~W/m$^2$ in good agreement with the
value of 0.829$\pm$0.017~W/m$^2$ given by the PMOD composite
\citep{froehlich-2006}.

Variations in the brightness of the Sun are also produced by the
continuously evolving granulation pattern.
Granules are signatures of convective cells with sizes ranging from a few
hundred to roughly 2000~km.
Their central parts, where hot material streams up from the interior to the
surface, appear brighter, whereas at the outer edges the cooled darker
material flows down, producing an RMS contrast of approximately 14\% in the
green part of the solar spectrum \citep{danilovic-et-al-2008}.
The typical lifetimes of granules are about 5 to 10~minutes
\citep{hirzberger-et-al-99a}.
Granulation dominates irradiance variations on time scales shorter than
approximately half a day and is not important for the Earth's climate
\citep{solanki-et-al-2003a}.

\subsection{The fundamentals of the SATIRE model}
\label{satire}

Here we describe the basic assumptions and ingredients of SATIRE, a set of
codes to 
calculate spectral solar irradiance from solar disc images or activity proxies
such as sunspot numbers
\citep{fligge-et-al-98a,krivova-et-al-2003a,krivova-et-al-2007a,wenzler-et-al-2006a}.
The assumption 
underlying SATIRE is that all changes in the solar
irradiance on time scales longer than hours are solely due
to changes in the solar surface magnetic flux as traced through
surface features, such as spots and faculae (see Sec.~\ref{sources} for
more detail).
To obtain the solar irradiance at a given time, $t$, and wavelength,
$\lambda$, it then suffices to derive filling factors, $f^{a}(t)$,
quantifying how much of the solar surface is covered by a given surface
feature $a$ 
(where $a$ denotes umbra, penumbra, faculae or network), and to
calculate the emergent radiative flux of the feature $F^{a}(\lambda)$.
The solar irradiance, $S(t,\lambda)$, can then be calculated by summing the
emergent radiative fluxes of the quiet (inactive) Sun, $F^{q}(\lambda)$,
and those of the surface features according to filling factors
\begin{equation}
S(t,\lambda) \ =
   \sum_a f^a (t) F^a(\lambda) \, + \left(1 - \sum_a f^a (t)\right) \, F^q(\lambda).
\label{eq1}
\end{equation}
%

This formalism can be extended to include positional information, in which
case the global filling factors $f^a$ are replaced by local filling factors
$\alpha^a_{i,j}$, where $i,j$ are indices referring to a particular pixel
in a solar image or magnetogram.
The radiative fluxes $F^{a}$ are replaced by intensities $I^{a}(\mu)$, where
$\mu = \cos\Theta$ and $\Theta$ is the heliocentric angle.
Then Eq.~(\ref{eq1}) needs to be replaced by 
\begin{equation}
S(t,\lambda) = \sum_a\sum_{i,j}\left[\alpha^a_{\mu(i,j)}(t)I^a_{\mu(i,j)}(\lambda) +
               \left(1-\alpha^a_{\mu(i,j)}(t)\right)I^q_{\mu(i,j)}(\lambda)\right].
\label{eq2}
\end{equation}
Note that the radiative fluxes are independent of time; their calculation 
is discussed in the following paragraphs. The filling factors represent 
the changing aspect of the solar surface, and it is their time dependence 
that produces the variability in the solar irradiance.
The filling factors can be obtained from a variety of data sets, depending on 
the time scales considered.
The type of data available also determines whether Eq.~(\ref{eq2}) is
employed (basically limited to the satellite era) or Eq.~(\ref{eq1}), for
the longer time scales.
In order to distinguish between models adjusted to 
different time scales and based on different kinds of input 
data used to derive the filling factors,
we call them SATIRE-S (Satellite period, magnetogram-based,
Sect.~\ref{short}), SATIRE-T (Telescope era, from the sunspot number,
Sect.~\ref{cent}) and SATIRE-M (Millennia time scales, based on
cosmogenic isotope data, Sect.~\ref{mill}).
Note, that positional information for the photospheric magnetic features is
only available in case of SATIRE-S, when magnetograms are used to derive the
filling factors.

The emergent intensities are calculated from a set of model atmospheres 
using ATLAS9 \citep{kurucz-93,kurucz-2005}. 
This yields relatively low-resolution spectra (1 to 2~nm in the NUV and optical), 
but has the advantage of covering the whole wavelength range contributing significantly 
to the TSI while remaining computationally manageable. As ATLAS9 uses an LTE approach 
and combines the opacity contributions due to lines into opacity distribution functions, 
the accuracy of the calculations decreases at shorter wavelength and for the strongest lines. 
For these wavelength regions, more reliable results can be obtained using a 
correlation analysis as described in \citet{krivova-et-al-2006a}. 

The model atmospheres and resulting emergent intensities are described in 
detail by \citet{unruh-et-al-99}. The model for the quiet Sun is 
based on the standard solar model of \citet{kurucz-92b}; the umbral and 
penumbral models are radiative equilibrium atmosphere models 
with effective temperatures of 4500~K and 5400~K, respectively. 
The facular model, finally, is a modified version of FAL~P from 
\citet{fontenla-et-al-93}. 

Solar irradiance reconstructions based on these models have 
been used successfully to calculate the spectral variability over
a large range of wavelengths (see Sec.~\ref{ssi_short}); 
they also match the measured TSI on solar-cycle timescales 
(Sec.~\ref{tsi_short}). 
Despite these successes, a number of issues remain. One of these is 
that the use of 1-D atmospheres leads to the facular contrast being
overestimated towards the solar limb. Another is the assumption that 
all small-scale magnetic features are assumed to display the same wavelength
and limb-dependent contrast behaviour that is scaled according to their magnetic 
flux.
This assumption is unlikely to hold, as suggested  by the different thermal
structure of network and active region flux tubes \citep{solanki-86} and the
magnetic flux dependence of facular/network contrast
\citep{topka-et-al-97,ortiz-et-al-2002}.
One way to address both of these issues is to use realistic simulations of
magneto-convection in order to represent the 3-D structure of the solar
atmosphere and to derive from them the emergent intensities as a function of
the magnetic flux.

\begin{figure}
   \resizebox{\hsize}{!}{\includegraphics{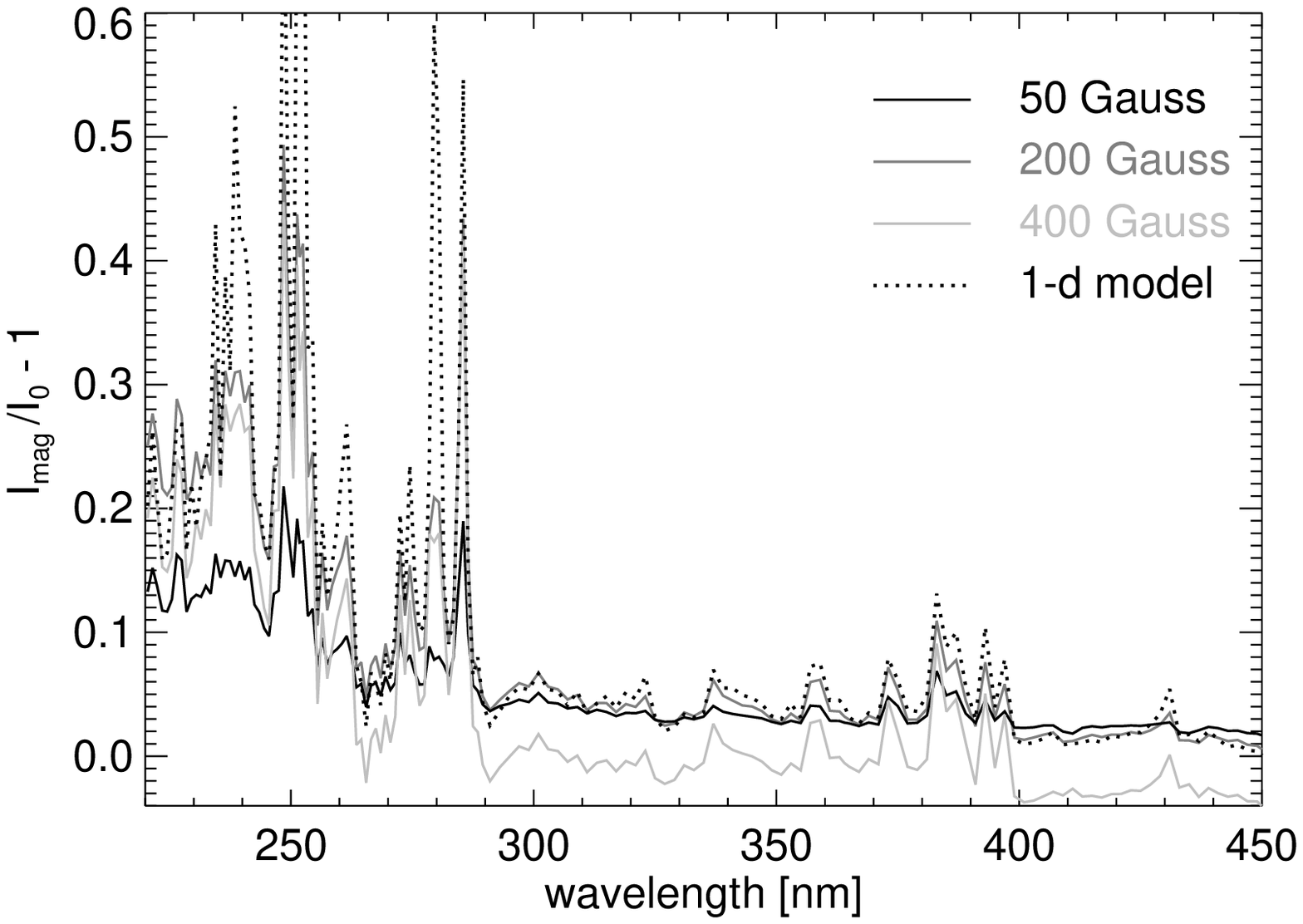}%
                         \includegraphics{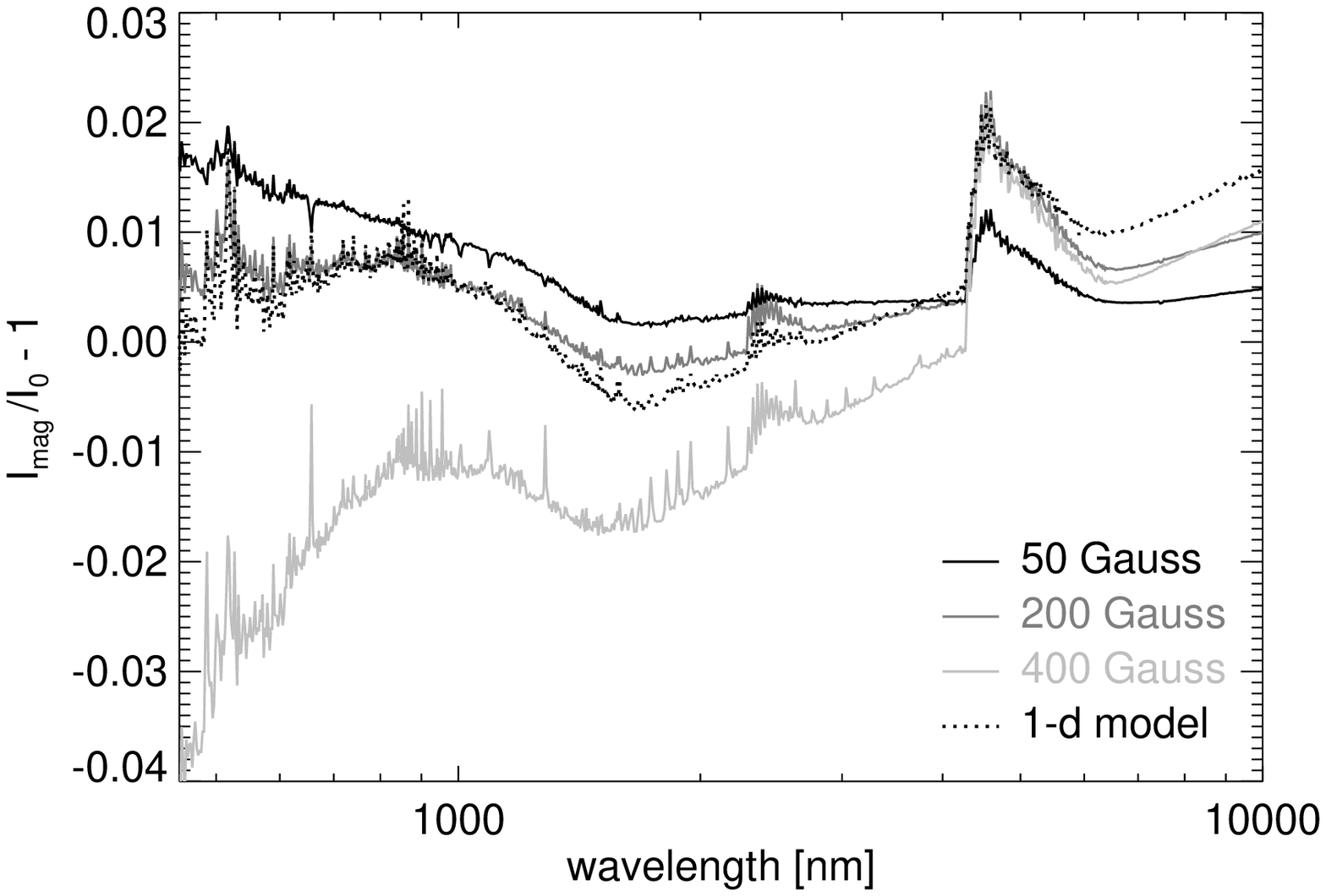}}
   \caption[]{Disk-centre contrasts of simulated convection with a mean magnetic 
	flux of 50~G, 200~G and 400~G compared to a non-magnetic model. The
	dotted line shows the contrast of the 1-D facular atmosphere
	used in the SATIRE reconstructions. This has been reduced by 30\% 
	for ease of comparison. }
   \label{fig:muram_contrasts}
\end{figure}

The first steps of
such an approach were presented by \citet{unruh-et-al-2009}, who used 
MURaM magneto-convection simulations \citep{voegler-et-al-2005} to calculate
disc-centre intensities. Figure~\ref{fig:muram_contrasts} shows first 
results for contrasts at $\mu=1$ (i.e. solar disc centre)
derived from magneto-convection simulations with mean magnetic 
fluxes of 50~G, 200~G and 400~G. Also shown is the contrast currently used 
in the SATIRE reconstructions (dotted line); it has been reduced by 
approximately 30\% so that it can be compared directly to the simulation results. 
A higher contrast is expected for the SATIRE 1-D model as this represents the 
brightest faculae, while the MURaM contrast is averaged over a computational 
box that contains a mix of magnetic flux tubes and field-free regions.

Overall, the spectral behaviour of the 200-G simulations and of the SATIRE contrasts 
is very similar. The 50-G simulations that are more indicative of the solar 
network show a slightly flatter spectral shape with a slower
increase of the contrast
at shorter wavelengths and less darkening in the near-infrared. 

\section{Solar Irradiance Over the Satellite Period (SATIRE-S)}
\label{short}

\subsection{Temporal Changes from Magnetograms}
\label{mags}

The most detailed information on the temporal evolution of the solar surface
magnetic features, needed in order to calculate irradiance variations, is
provided by full disc magnetograms.
Since 1996 such magnetograms have been regularly recorded by the Michelson
Doppler Imager (MDI) on SoHO \citep{scherrer-et-al-95}.
Ground-based magnetograms with a sufficiently large spatial resolution,
e.g., from the Kitt Peak National Solar
Observatory (NSO KP) are available back to 1974
\citep{livingston-et-al-76,jones-et-al-92}, although the temporal
coverage and the quality become significantly poorer for the earlier period.

Magnetograms and continuum images recorded as close in time as possible are
used to identify different magnetic features on the solar surface: sunspot
umbrae, sunspot penumbrae, faculae and the network, and to determine which
fraction of the solar surface is occupied by each component at a given
$\mu$, i.e. the filling factors $\alpha$ (see Sect.~\ref{satire}).
Spot umbrae and penumbrae are identified by their continuum contrasts.
Pixels belonging to spots are then excluded from the (co-aligned)
magnetograms and the remaining magnetogram signal is considered to be
faculae and the network.
Pixels with the magnetic signal below the noise threshold are counted as
quiet Sun.

The intensity spectra of all components are then weighted by the
corresponding filling factors and summed up to give the irradiance as a
function of wavelength using Eq.~(\ref{eq2}).
Integration over all wavelengths results in total solar irradiance at
time $t$.
The procedure is repeated for every day when a magnetogram and a continuum
image are available.
The models employing MDI and NSO KP data are described in detail by
\citet{krivova-et-al-2003a} and
\citet{wenzler-et-al-2004a,wenzler-et-al-2005a,wenzler-et-al-2006a},
respectively.
\citet{wenzler-et-al-2004a} also compared the reconstructions based on these
2 sets of magnetograms to each other.

The model has a single free parameter, $B_{\rm sat}$, the value of the
magnetogram signal at which magnetic field in the solar photosphere fills
the whole magnetogram pixel $(i,j)$.
$B_{\rm sat}$ is found from the best agreement with the TSI measurements.
Note that one aim of the recently started work on computing the spectra from
MHD simulations (see Sect.~\ref{satire}) is to do away with this free
parameter.

\subsection{Total Solar Irradiance}
\label{tsi_short}

\citet{krivova-et-al-2003a} employed MDI magnetograms to
reconstruct the TSI between 1996 and 2002.
Excellent agreement between the PMOD TSI composite and the model was
obtained with a correlation coefficient of $r_c=0.96$ ($r_c^2=0.92$).
\citet{wenzler-et-al-2004a,wenzler-et-al-2005a,wenzler-et-al-2006a}
have extended this reconstruction back to 1974 using NSO KP magnetograms and
continuum images.
The correlation for the whole period 1974--2003 is somewhat lower than when
MDI data are used ($r_c=0.91$, $r_c^2=0.83$), which is not surprising due to
the significantly lower quality and artefacts seen in ground-based data
(especially before 1992 when an older instrument was used at the NSO KP)
compared to those taken from space.
Nevertheless, the good agreement between the data and the model provides
strong support for the model's basic assumption, namely
 that the dominant part of the irradiance variations on time scales longer
than about a day is caused by the evolution of solar surface magnetic
features.

\begin{figure}
   \resizebox{\hsize}{!}{
   \includegraphics{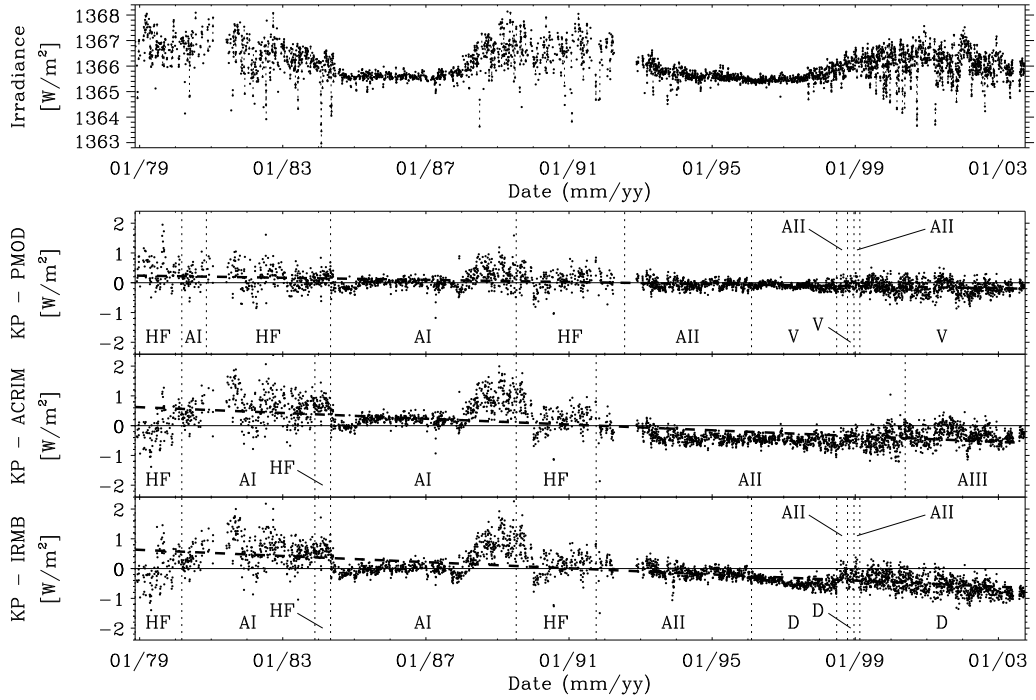}}
   \caption[]{The \emph{top panel} shows the reconstructed daily total solar
 irradiance based on NSO KP data for 3528 days between
 1978 and 2003, i.e. from the late rising phase of cycle 21 to the declining
 phase of cycle 23. The three \emph{lower panels} show the difference
 between the reconstructed TSI and the PMOD, ACRIM and IRMB composites,
 respectively. Each dot represents a daily value. The horizontal solid
 lines indicate a difference~=~0, the thick dashed lines are linear
 regressions. The dotted vertical lines indicate periods when the
 individual data sets from HF, AI, AII \& AIII (ACRIM I, II \& III), V
 (VIRGO: Variability of solar IRradiance and Gravity Oscillations
 experiment on SoHO) and D (DIARAD: DIfferential Absolute RADiometer) were
 used for the composites.
From Wenzler et al. (2009)}\nocite{wenzler-et-al-2009a}
   \label{fig:tsi_thomas}
\end{figure}

\citet{wenzler-et-al-2009a} have also compared the model to all three
currently available TSI composites: PMOD, ACRIM and IRMB.
For this, they have varied the value of the single free parameter ($B_{\rm
sat}$, see Sect.~\ref{mags}) in order to find the best fit to each of the
composites individually.
The overall best fit is found with the PMOD composite ($r_c=0.91$, slope
0.98), whereas the agreement with the other two composites is markedly
poorer ($r_c=0.79$, slope 0.82 for ACRIM and $r_c=0.82$, slope 0.81 for
IRMB).
This can be judged from Fig.~\ref{fig:tsi_thomas}, which displays the
difference between the TSI reconstructed by SATIRE-S and each of the three
composites.

\citet{scafetta-willson-2009} proposed that these reconstructions can be
used to bridge the so-called ACRIM gap (see Sect.~\ref{obs_tsi}) and to
create a `mixed' ACRIM-1~-- SATIRE~-- ACRIM-2 composite.
They have compared ACRIM-1 and ACRIM-2 data directly to the model, in order
to cross-calibrate the data from the two instruments.
These authors have, however, used the SATIRE-T model described in
Sect.~\ref{cent}, which is not suited for such an analysis.
As discussed later, SATIRE-T is based on the historic sunspot number
record instead of on magnetograms and continuum images, so that it is based
on Eq.~(\ref{eq1}) rather than  the more realistic Eq.~(\ref{eq2}).
It is therefore significantly less accurate than SATIRE-S on time
scales of weeks to several months.
These are, however, the most critical time scales for such a comparison of the
data and the model.
This indicates that caution needs to be exercised when considering the
results of \citet{scafetta-willson-2009}, in particular their conclusion
about the upward trend between the minima in 1986 and 1996.

\begin{figure}
   \resizebox{\hsize}{!}{
   \includegraphics{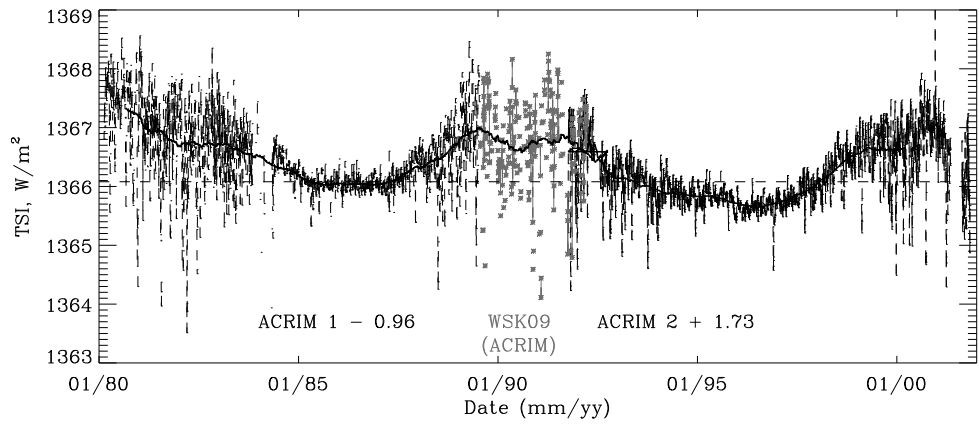}}
   \caption[]{'Mixed' TSI composite constructed from ARCIM-1 and ACRIM-2
data (black dashed line), with the gap bridged using the SATIRE-S (WSK09
ACRIM) model
(asterisks connected by grey solid line when there are no gaps).
The heavy solid line is the 1-year smoothed TSI, and the horizontal dashed
line shows the level of the minimum preceding cycle 22.
From \citet{krivova-et-al-2009b}.}
   \label{fig:scaf}
\end{figure}

\citet{krivova-et-al-2009b} have repeated this analysis employing the
more appropriate SATIRE-S model.
The `mixed' ACRIM-1~-- SATIRE-S~-- ACRIM-2 composite is shown in
Fig.~\ref{fig:scaf}.
It shows no increase in the TSI from 1986 to 1996, in contrast to the ACRIM
composite.
Actually, a slight decrease is found.
The magnitude of this decrease cannot be estimated very accurately from such
an analysis but it lies between approximately 0.2 and 0.7~W/m$^2$
(0.013--0.05\%).

\subsection{Spectral Solar Irradiance}
\label{ssi_short}

Early comparisons between SATIRE-S spectral irradiance reconstructions 
and UARS/SUSIM measurements were presented by \citet{krivova-solanki-2005a}
and \citet{krivova-et-al-2006a} for wavelengths below 400~nm. These comparisons
showed that there is good agreement between the measured and modelled 
wavelength integrated flux between 220 and 240~nm. 

\citet{krivova-et-al-2006a} then
used this wavelength band as a reference to derive regression coefficients
measuring the response of those wavelength regions that are not well 
represented by the ATLAS9 fluxes. In this way SATIRE can be extended 
down to 115~nm. For wavelengths above 270~nm, the sensitivity of UARS/SUSIM 
is not sufficient to measure solar variability reliably, so that modelled 
fluxes (i.e. those based on model atmospheres) are currently more reliable.

\begin{figure}
   \resizebox{\hsize}{!}{\includegraphics{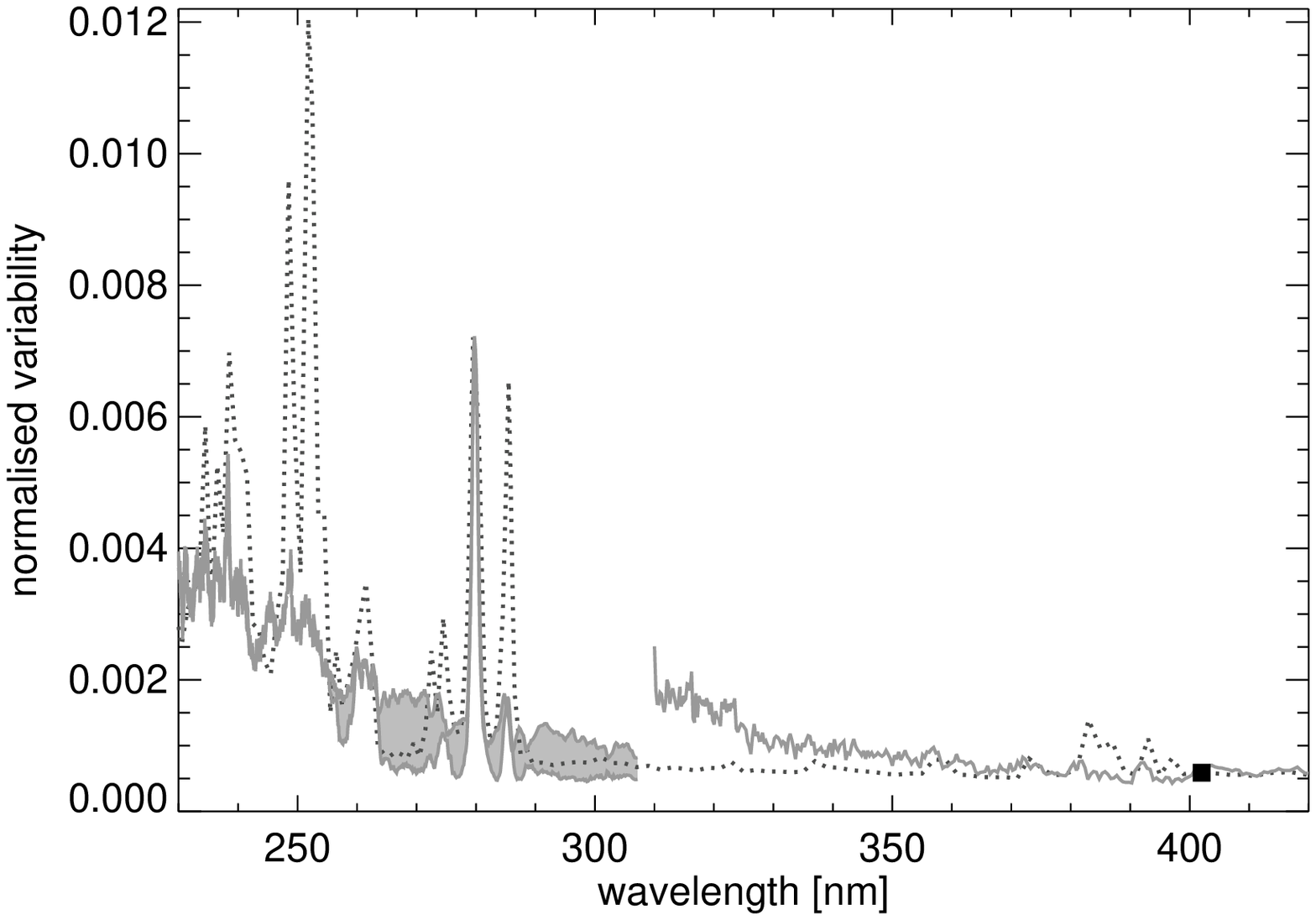}%
                        \includegraphics{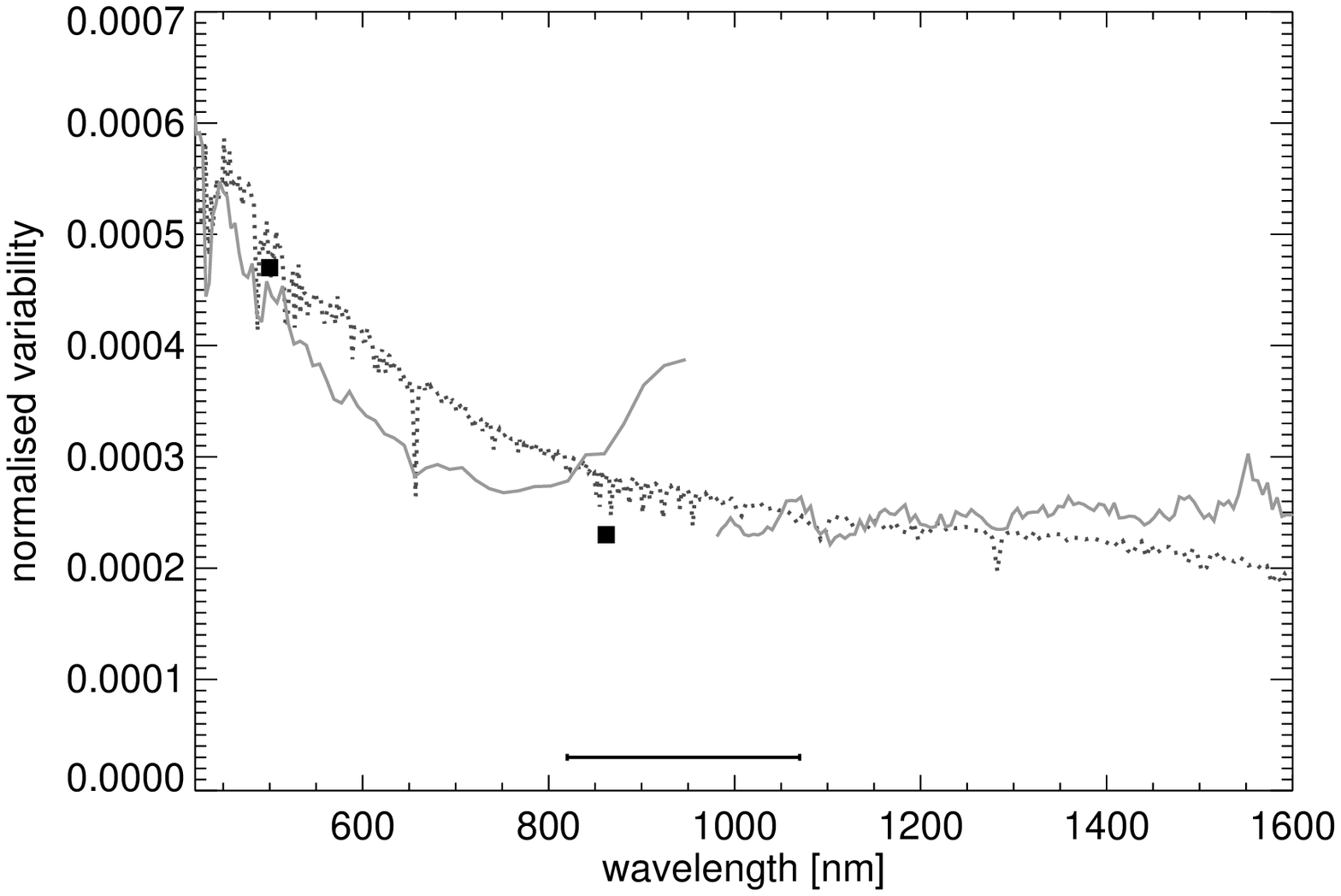}}
   \caption[]{Normalised variability from May 2004 to July 2004 as measured with 
	SORCE/SIM (solid lines) and modelled with SATIRE (dotted lines) for wavelengths
	spanning 230~nm to 420~nm (left-hand side) and 420~nm to 1600~nm
	(right-hand side). The filled squares indicate the variability measured 
	in the three SoHO/VIRGO SPM bands at 402~nm, 501~nm and 863~nm.}
   \label{fig:sorce_comparison}
\end{figure}

\begin{figure}
   \resizebox{\hsize}{!}{\includegraphics{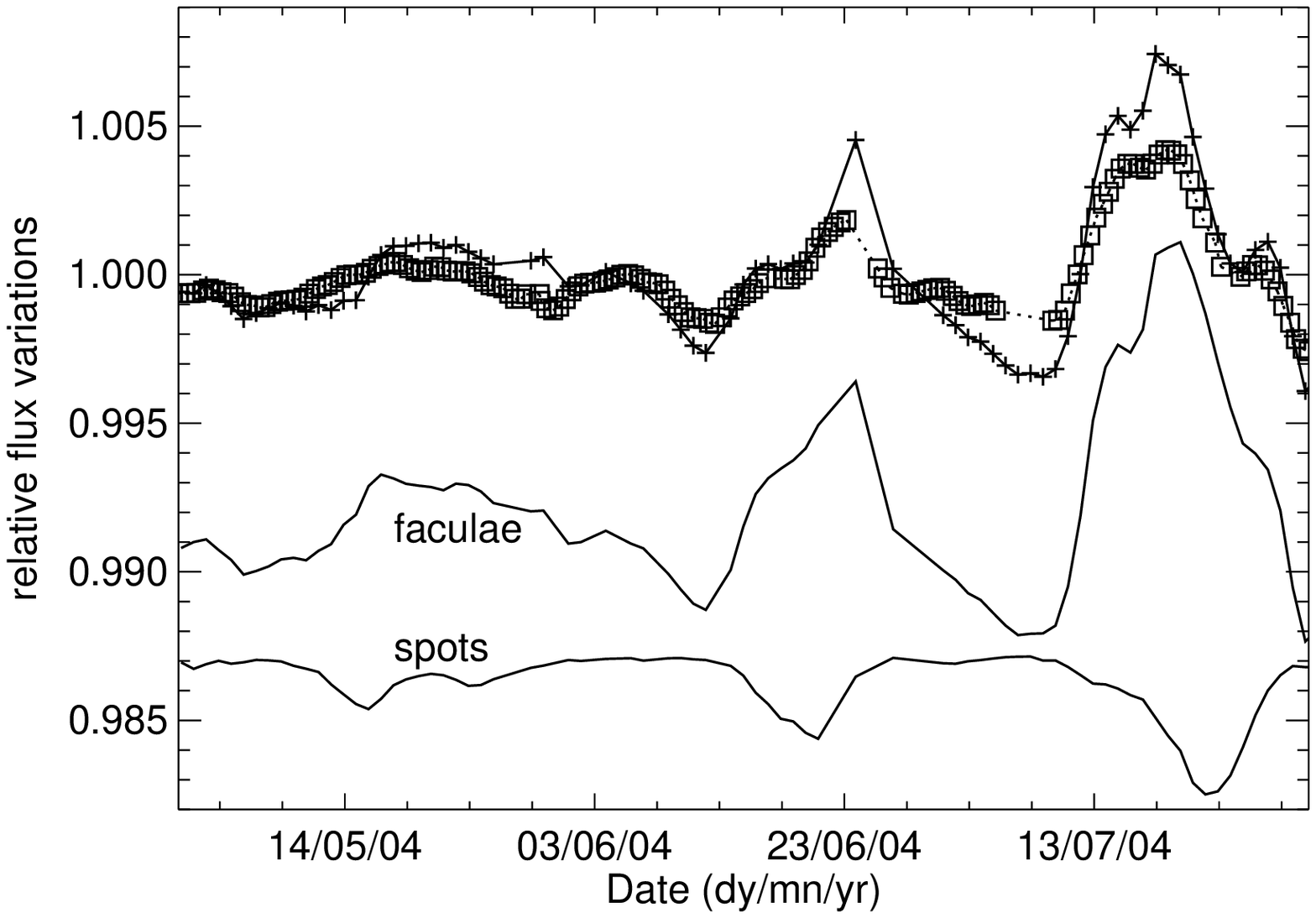}%
                        \includegraphics{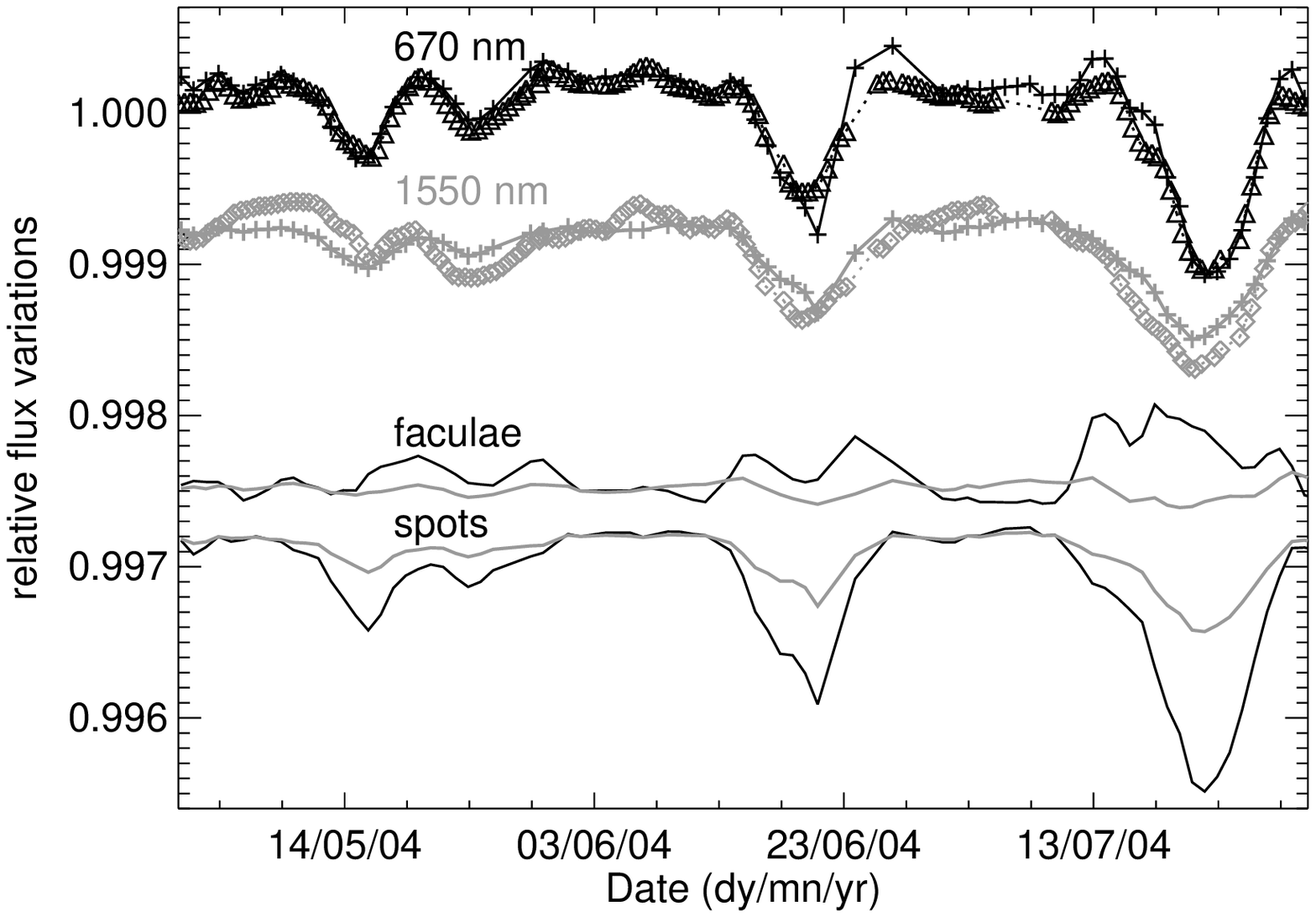}}
   \caption[]{{\em Top curves:} flux variations for the Mg~II band ({\em
left-hand side}) and for two bands centred at 670 and at 1550~nm
({\em right-hand panels}).
The dotted lines linking the squares and triangles represent SIM data, the
plus signs linked by the solid lines show the model calculations.
{\em Bottom curves:} modelled times series for the facular and spot
contributions.
The upper lines are for the faculae, the lower for the spots.
Following \citet{unruh-et-al-2008}.
}
   \label{fig:sorce_lc}
\end{figure}

Comparisons in the visible and near-infrared have only become possible
with the launch of SORCE and ENVISAT (Sect.~\ref{obs_ssi}). 
First results from SORCE/SIM were presented by \citet{fontenla-et-al-2004}. 
Comparisons between SIM data and SATIRE-S covering 3 solar rotations in 2004
were carried out by \citet{unruh-et-al-2008}. The measured and modelled
variability is shown in Fig.~\ref{fig:sorce_comparison}. There is good
agreement between SATIRE-S and SORCE/SIM, in particular for wavelengths
above 360~nm. Between 310~nm and 360~nm detailed comparisons are hampered as
the sensitivity of the detector is not quite sufficient to measure solar
variability adequately and the modelled irradiances yield more reliable
results.
For wavelengths above 550~nm SATIRE-S shows a slightly flatter
decrease in variability than the SORCE measurements, though detailed
light-curve comparisons show excellent agreement in the short-term
variability.

Figure~\ref{fig:sorce_lc} shows measured and modelled time series for the
Mg~II band (left-hand side) and for two bands centred at 670 and at
1550~nm, respectively (right-hand panels).
The upper curves in the plots show the measured and modelled solar
variability while the bottom curves show the modelled contributions of the
faculae and spots to the irradiance variations.
The figures illustrate the change in the lightcurve aspects as one moves
from the UV to the near-IR.
In the UV, as illustrated by the Mg~II (at around 280~nm) band, the
influence of the spots is so small that their darkening effect is more than
compensated for by the faculae, even on solar rotational time scales.
Furthermore, the facular contrast increase towards the limb is not
sufficient to counteract the projection effects and the Sun appears
brightest when the spot groups are at disc centre.
At longer wavelengths (e.g., at 670~nm), the facular brightenings are much
weaker and show a double-peaked structure.
Thus faculae produce most of the brightening when near the limb and very
little, if any, when at disc centre.
In the near-IR (as illustrated by the 1550~nm band), only very little facular
brightening is seen close to the limb, which is compensated by the earlier
onset and longer duration of spot darkening.
The overall effect of faculae at 1550~nm is one of darkening.

\begin{figure}
   \resizebox{\hsize}{!}{
   \includegraphics{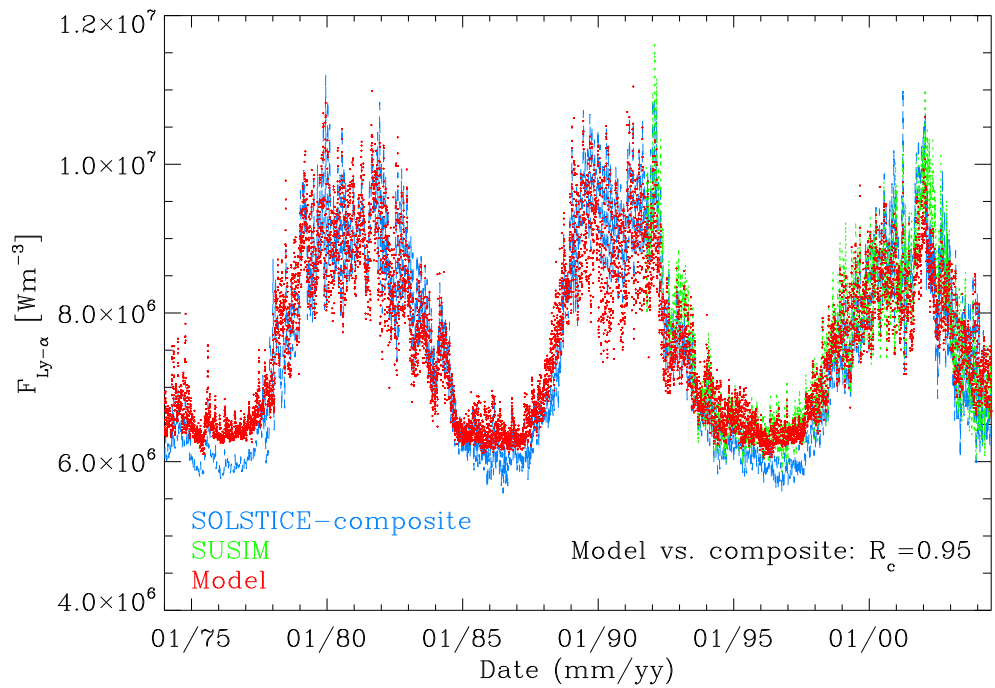}}
   \caption[]{Solar Ly-$\alpha$ irradiance since 1974: reconstructed by
SATIRE-S (light grey crosses; in online version red), measured by the SUSIM
instrument (grey dashed line; in online version green) and compiled by
Woods et al. (2000; black dotted line; in online version blue). From Krivova
et al. (2009).}
\nocite{woods-et-al-2000a,krivova-et-al-2009a}
   \label{fig:lya}
\end{figure}

\citet{krivova-et-al-2009a} have reconstructed solar UV irradiance in the
range 115--400~nm over the period 1974--2007 by making use of the empirical
extension of the SATIRE models by \citet{krivova-et-al-2006a} employing
SUSIM data.
The evolution of the solar photospheric magnetic flux in this case was
described by the magnetograms from the KP NSO between 1974 and 2003 and by
the MDI instrument on SoHO since 1996.
The gaps in daily data were filled using the Mg~II core-to-wing ratio and
the solar F10.7~cm radio flux.

Figure~\ref{fig:lya} shows the reconstructed solar Ly-$\alpha$ irradiance
(light grey crosses; red in online version), SUSIM measurements (grey dashed
line / green) and the composite time series (black dotted line / blue)
compiled by \citet{woods-et-al-2000a}.
The latter record includes measurements from the Atmospheric Explorer E
(AE-E, 1977--1980), the Solar Mesosphere Explorer (SME, 1981--1989), UARS
SOLSTICE (1991--2001) and the Solar EUV Experiment (SEE) on the TIMED mission
launched in 2001.
The gaps are filled in using proxy models based on Mg~II and F10.7 indices.
The F10.7-based model is also used for the period before 1977.
All measurements and models are adjusted to the SOLSTICE absolute values.
The model agrees well with the SUSIM data, which confirms that the
semi-empirical technique works well.
There is some difference in the magnitude of the Ly-$\alpha$ solar cycle
variation between SOLSTICE and SUSIM, which remains also if the model is
compared to the composite.
This is because the model is adjusted to SUSIM by construction.
Other than that the model agrees with the composite record very well with a
correlation coefficient of 0.95.

\begin{figure}
   \resizebox{\hsize}{!}{
   \includegraphics{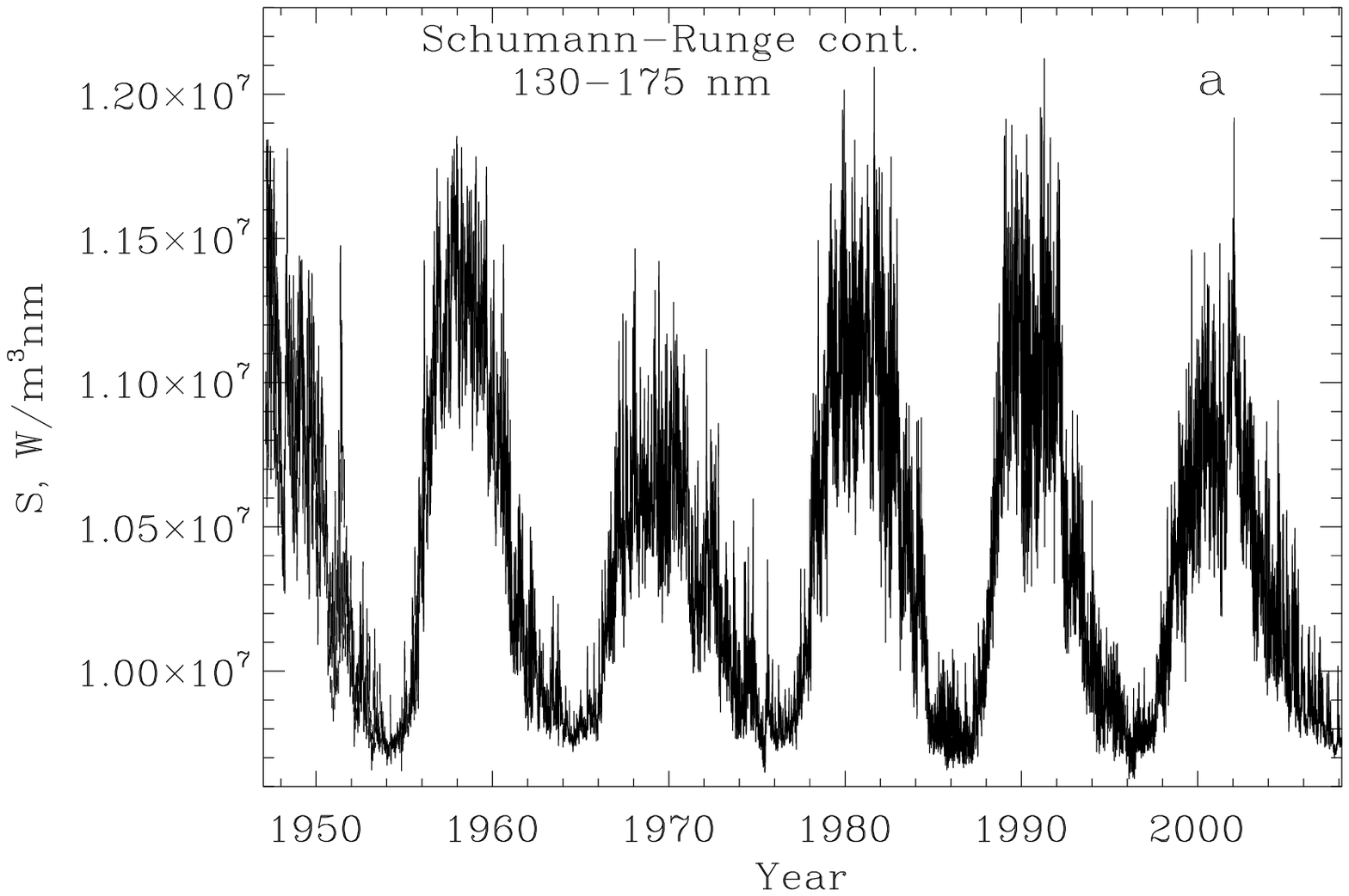}%
   \includegraphics{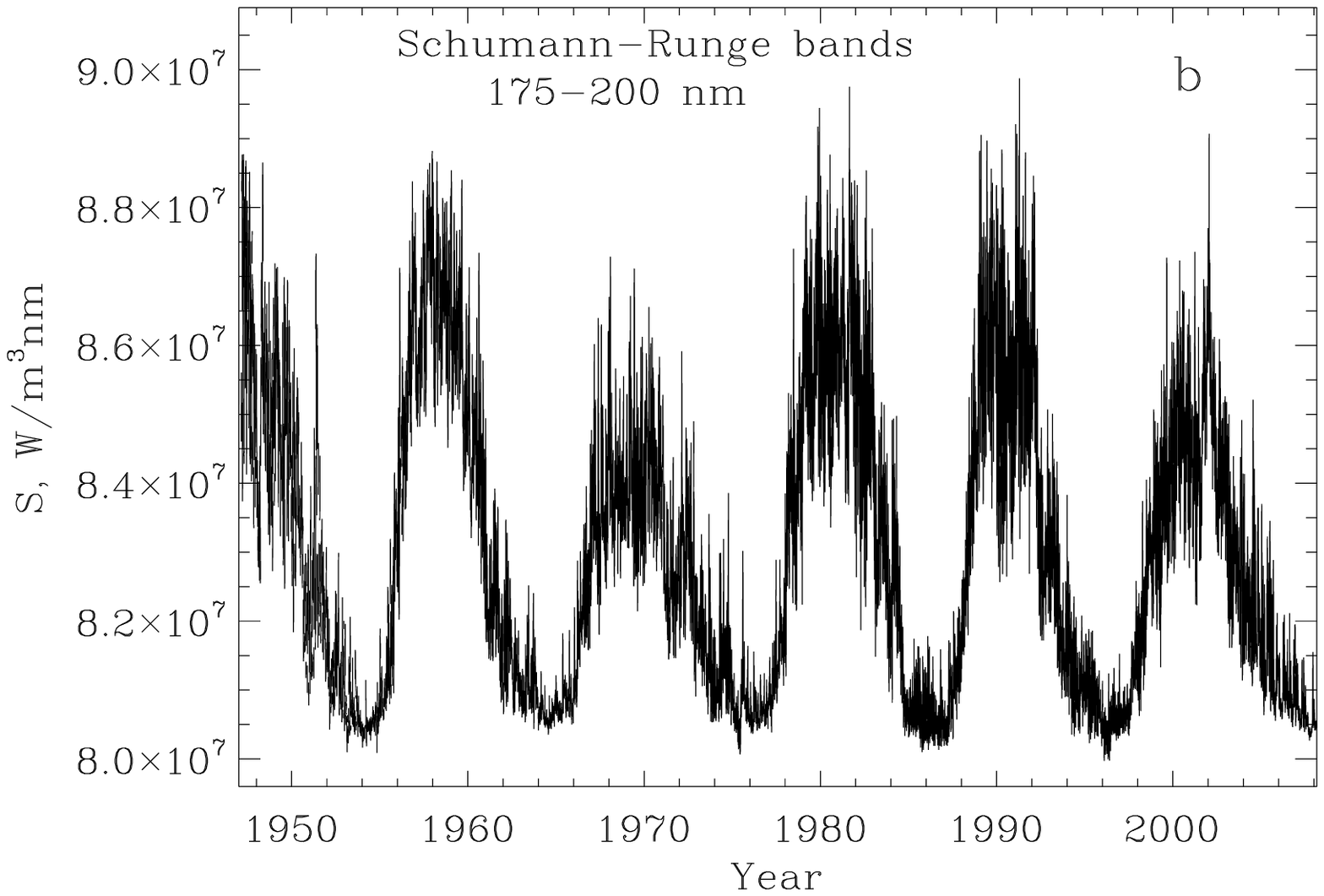}}
   \resizebox{\hsize}{!}{
   \includegraphics{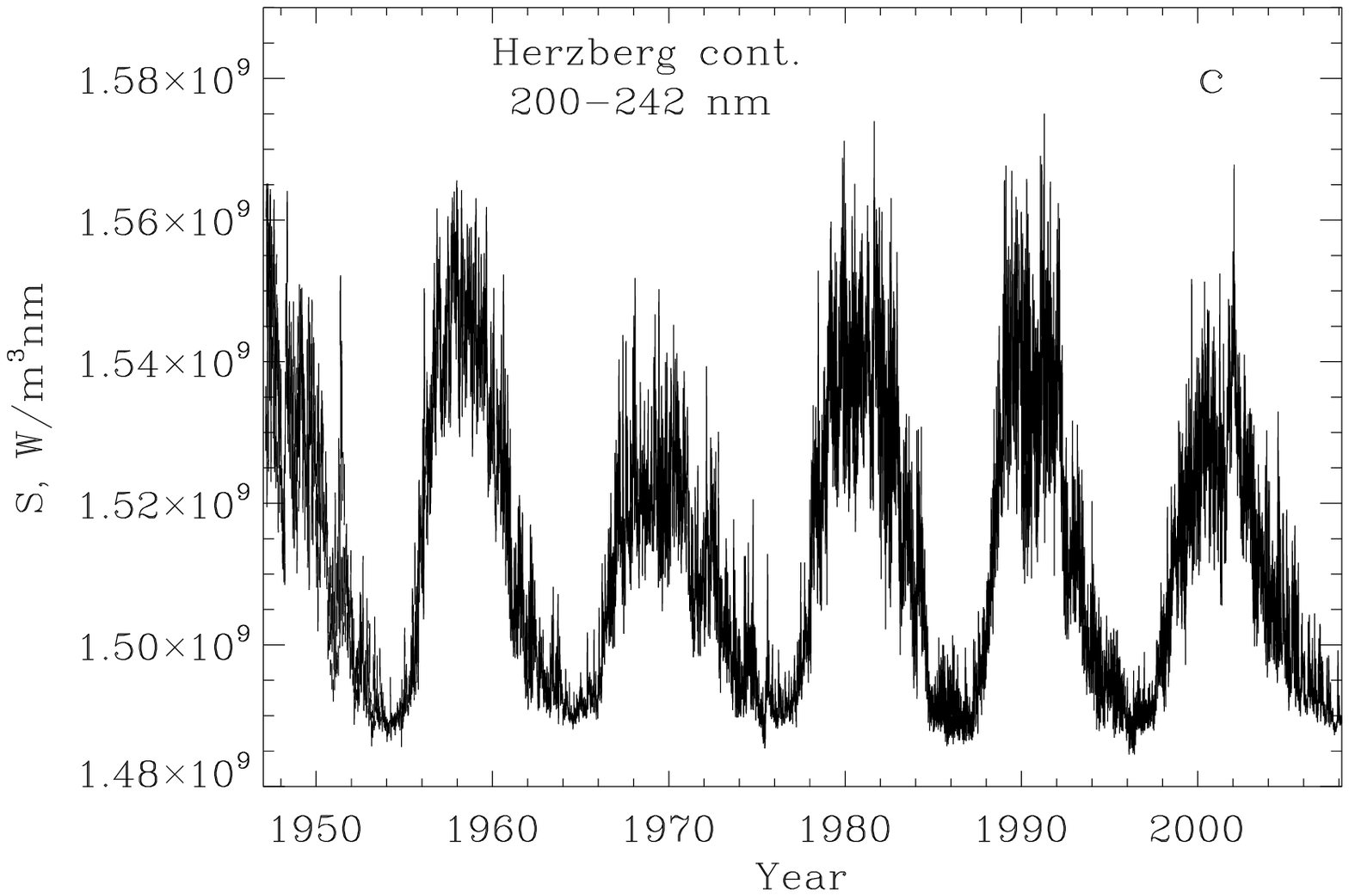}
   \includegraphics{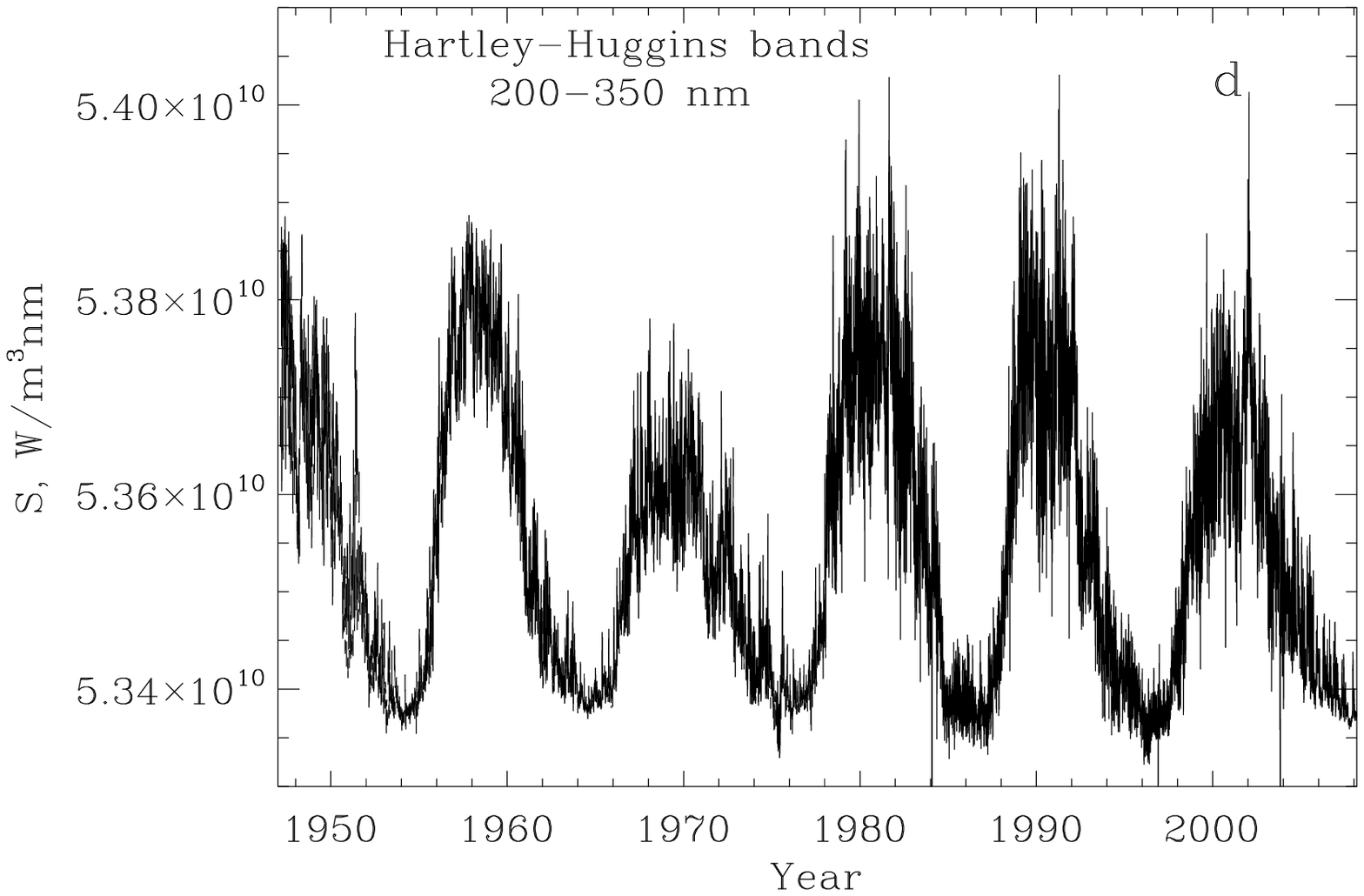}}
   \resizebox{\hsize}{!}{
   \includegraphics{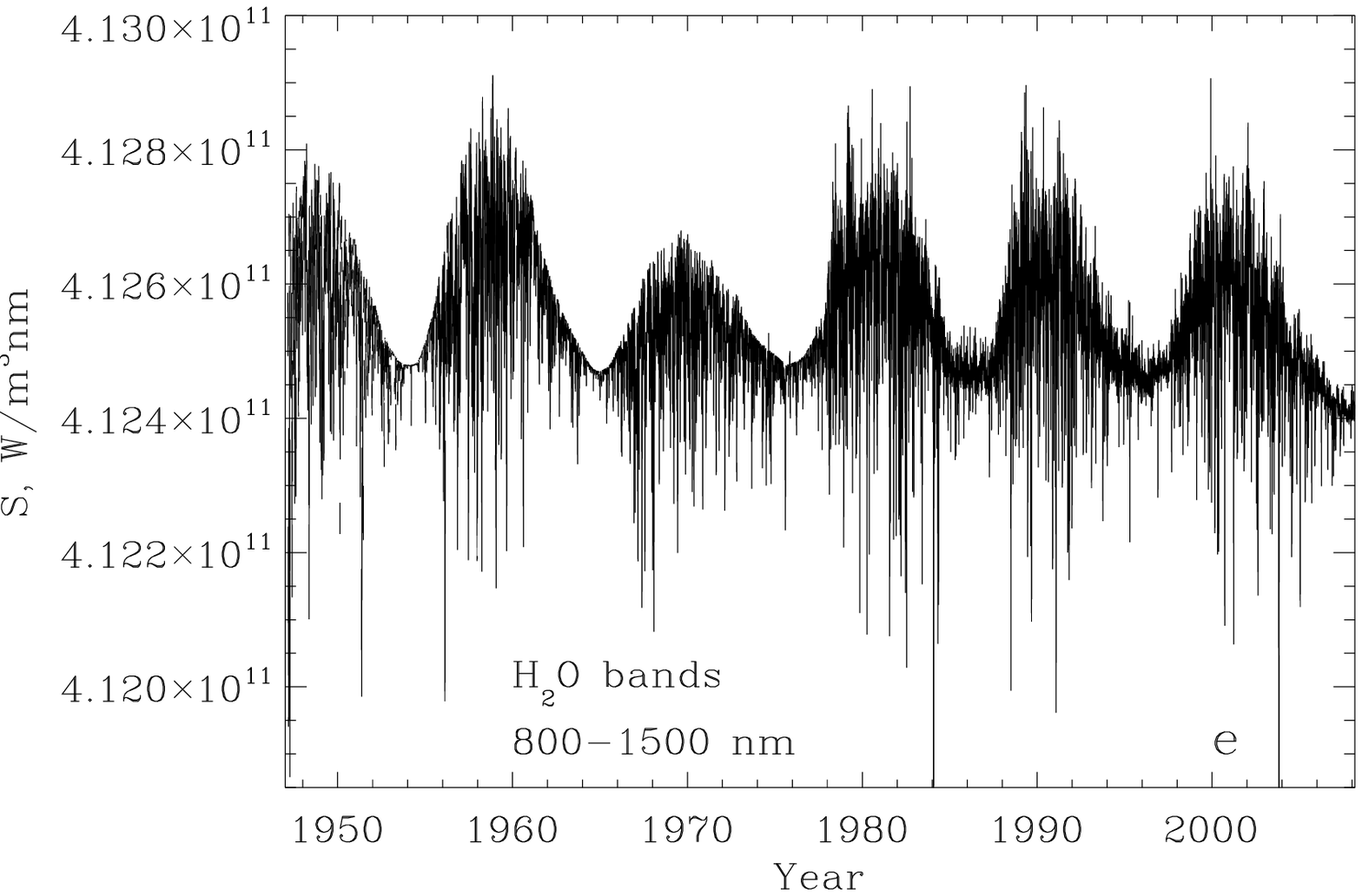}
   \includegraphics{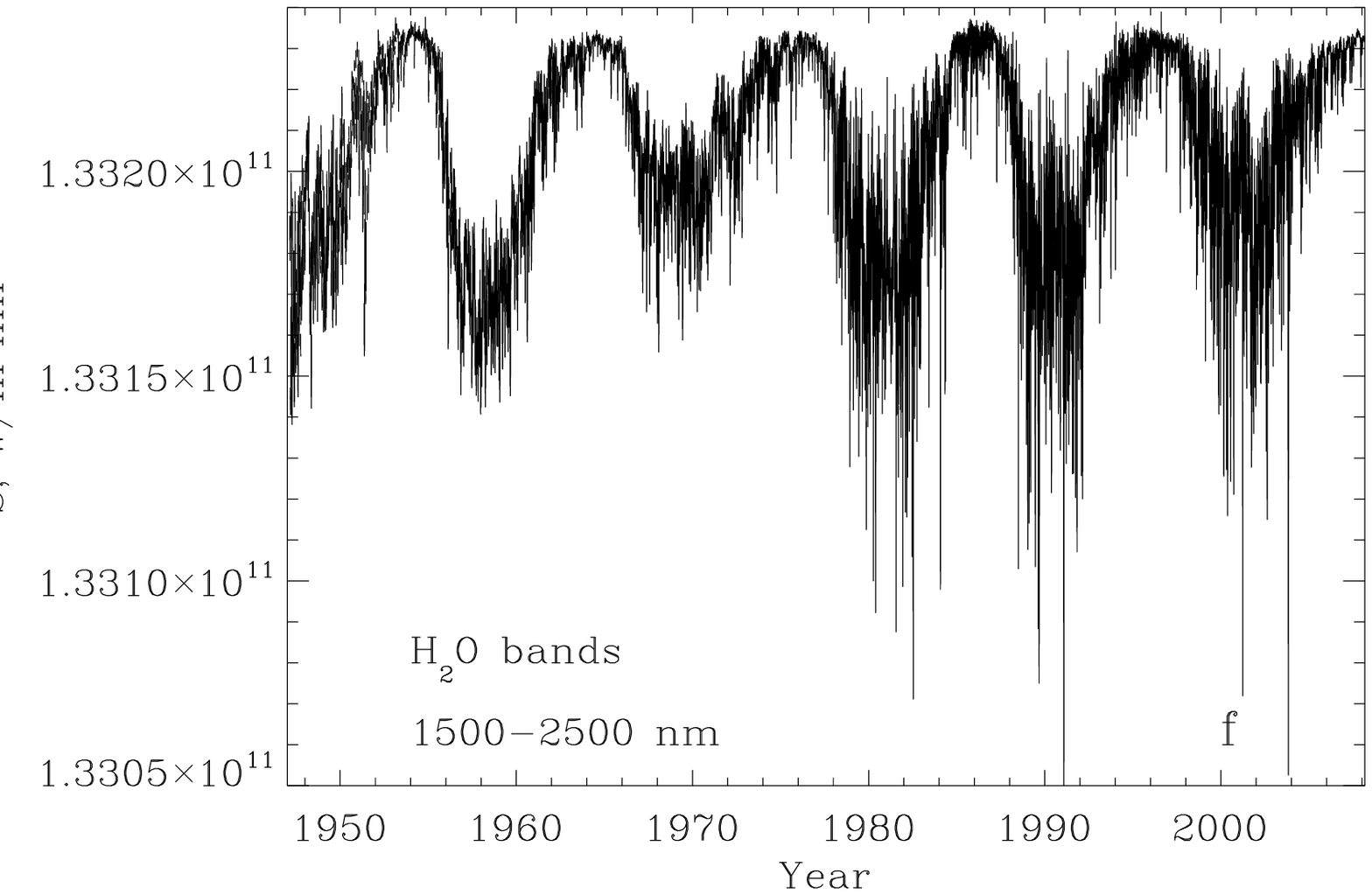}}
   \caption[]{Reconstructed solar irradiance in the period 1947--2007
integrated over the wavelength ranges: (a) 130--175~nm (Schumann-Runge
continuum), (b) 175--200~nm (Schumann-Runge bands), (c) 200--242~nm (Herzberg
continuum), (d) 200--350~nm (Hartley-Huggins bands), (e) 800--1500~nm
(H$_2$O bands) and (f) 1500--2500~nm (H$_2$O bands).
}
   \label{fig:bands}
\end{figure}

Employing the solar F10.7~cm radio flux data \citep{tanaka-et-al-73}, it is
possible to extend the reconstruction back to 1947.
For this, the radio flux measurements are used to calculate the irradiance
at 220--240~nm, in the same way as done by \citet{krivova-et-al-2009a}, who
employed these data in order to fill in the gaps in their reconstructions for
the period 1974--2007.
Irradiance at other
wavelengths is then calculated from the irradiance in
the reference range, 220--240~nm, by applying the technique of
\citet{krivova-et-al-2006a} described above.
Figure~\ref{fig:bands} shows the reconstructed solar irradiance
integrated over spectral ranges of particular interest for climate studies
as a function of time: (a) 130--175~nm, (b) 175--200~nm, (c) 200--242~nm,
(d) 200--350~nm, (e) 800--1500~nm and (f) 1500--2500~nm.
The model is based on radio flux data before 1974 and on magnetograms after
1974.
Solar UV irradiance (panels a--d) varied
by a factor 10 to 100 more than the TSI.
Interestingly, the irradiance in the range between around 1500~nm and
2500~nm (panel f)
shows an inverse solar cycle variation compared to the TSI and
other spectral ranges.
This inverse variation was also noticed by \citet{harder-et-al-2009} from
the analysis of SORCE data.
This is due to the low or even negative contrast of faculae at these
wavelengths \citep[][see also Fig.~\ref{fig:sorce_lc}]{unruh-et-al-2008}, so that
their brightening (if any) no longer compensates darkening due to sunspots.

\section{Solar Irradiance Since the Maunder Minimum (SATIRE-T)}
\label{cent}

\subsection{Temporal Changes from the Sunspot Number}
\label{mfmodel}

The advantage offered by high resolution magnetograms of having information
on the spatial distribution of the solar surface magnetic features as well
as the magnetic field strength is not available
prior to 1974.
Models going further back in time have to content themselves with sparser
data.
One such historic record is the sunspot number.
It is widely used for reconstructions of solar irradiance on time scales
longer than a few decades and is, in fact, the only direct proxy of solar
magnetic activity going back to the Maunder minimum (a period of unusually low
solar activity in the 17th century).

\citet{solanki-et-al-2000,solanki-et-al-2002} have constructed a
physical model which allows a reconstruction of the evolution of the
solar photospheric magnetic field from the sunspot number.
In this way, it is possible to calculate not just the total magnetic flux
emerging on the solar surface at a given time, but also individual
contributions of different components: active regions (sunspots and
faculae), ephemeral regions (smaller short-lived bipolar magnetic regions)
and the open flux (a small part of the photospheric magnetic flux whose
field lines reach into the heliosphere).
The calculated open magnetic flux is found to be in good agreement with the
empirical reconstruction of the heliospheric magnetic field since 1868 by
\citet{lockwood-et-al-99,lockwood-et-al-2009} from the geomagnetic
$aa$-index.
Also, the modelled total photospheric magnetic field reproduces the
measurements since 1974 carried out at the Mt. Wilson, Kitt Peak and Wilcox
Solar Observatories \citep{arge-et-al-2002,wenzler-et-al-2006a}.

\citet{krivova-et-al-2007a} and \citet{balmaceda-et-al-2007} made use of
this model in order to also reconstruct solar total irradiance since 1700
and 1610, respectively.
In order to calculate the filling factors (Sect.~\ref{satire}) of sunspots as
a function of time, they employed the sunspot areas.
These have been measured since 1874 \citep[see][]{balmaceda-et-al-2009} and
were derived for the earlier period using a linear relationship between
sunspot areas and numbers.
The sunspot contribution is then subtracted from the active region magnetic
flux to assess the flux present in the form of faculae.
The sum of the ephemeral region and open magnetic flux describes the
evolution of the network.

The network and facular magnetic fluxes are then converted into
corresponding filling factors (see Sect.~\ref{satire}) assuming that the
filling factors are proportional to the corresponding magnetic field
strengths until the saturation limit ($B_{sat}$, Sect.~\ref{mags}) is
reached.
Finally, the irradiance is calculated by summing up disc-integrated
brightnesses (i.e. radiative fluxes) of different components
(the quiet Sun, sunspot umbrae, sunspot penumbrae,
faculae and the network) weighted by
their filling factors as described in Sect.~\ref{satire}.
In addition to the heliospheric flux reconstruction by
\citet{lockwood-et-al-99} and the total magnetic flux data, the model
also reproduces well the measured TSI since 1978, although the accuracy of
the model is reduced on time scales of weeks to about a year
\citep[][see Sect.~\ref{tsi_short}]{krivova-et-al-2009b}.

\subsection{Irradiance}
\label{irr_cent}

\begin{figure}
   \resizebox{\hsize}{!}{
   \includegraphics{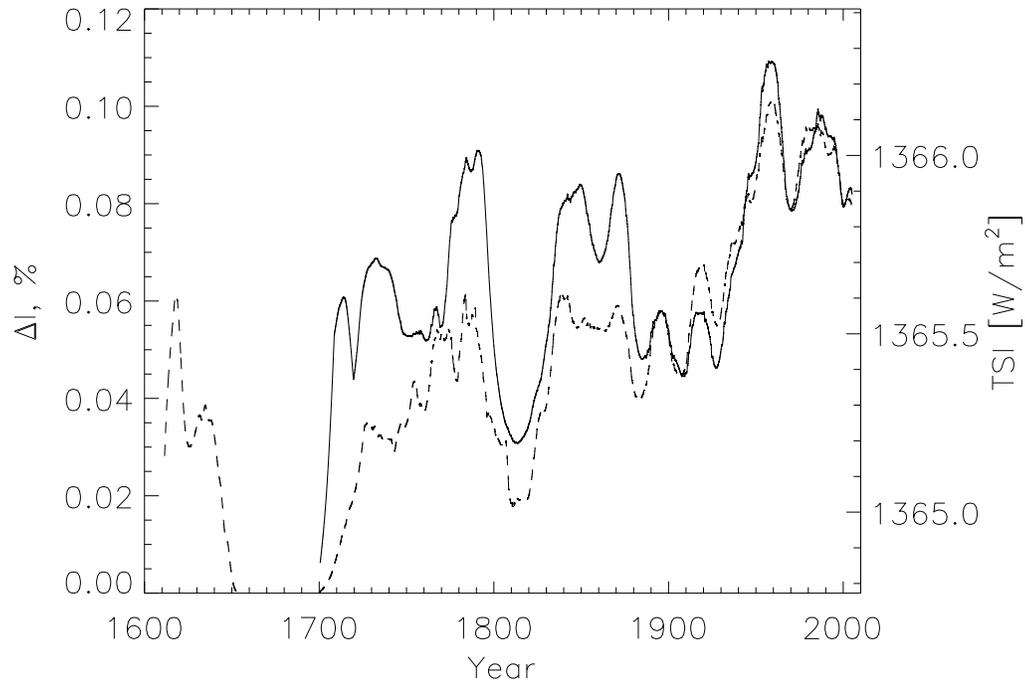}}
   \caption[]{11-yr running mean of the reconstructed TSI based on the
             Zurich (solid line) and Group sunspot number (dashed line).
}
   \label{fig:tsimm}
\end{figure}

\citet{krivova-et-al-2007a} have employed the Zurich sunspot number in order
to reconstruct solar total irradiance since 1700, whereas
\citet{balmaceda-et-al-2007} used the Group sunspot number in order to go back
to 1610.
The two reconstructions are shown in Fig.~\ref{fig:tsimm}.
They show slightly different irradiance levels during the 18th and 19th
centuries, but both suggest that, averaged over 11~years, the TSI has
increased by about 1.3~Wm$^{-2}$ since the end of the Maunder minimum.
By applying different assumptions regarding the evolution of the ephemeral
region magnetic flux, \citet{krivova-et-al-2007a} have also estimated the
uncertainty range as 0.9 to 1.5~Wm$^{-2}$.
The lower limit is in good agreement with the results of the modelling
effort by \citet{wang-et-al-2005}, who neglected the extended length of the
ephemeral region cycle, which is the main source of the secular trend in the
SATIRE-T model.
The upper limit of 1.5~Wm$^{-2}$ is close to the value of 1.7~Wm$^{-2}$ from
\citet{foster-2004}, cf. \citet{lockwood-2005}, who assumed that the network
disappeared completely during the Maunder minimum.
A comparably low value for the magnitude of the secular increase has also
been obtained by \citet{crouch-et-al-2008} from their model of TSI
variations.
The model does not, however, go back to the Maunder minimum, but stops at
1874.

All recent estimates of the magnitude of the secular trend are thus
significantly lower than the values deduced earlier from stellar data
\citep{baliunas-jastrow-90,white-et-al-92,lean-et-al-95}.
Nevertheless, even though all recent models agree that TSI increased by
significantly less than 2~Wm$^{-2}$  since the Maunder minimum, considerable
uncertainty remains.
This is because all models have free parameters, or make assumptions that
cannot be rigorously tested.

A potential independent source of information on the magnitude of the
secular trend are the historic archives of full-disc solar images in
the Ca~II~K line.
Such observations were carried out by a number of observatories around the
globe since the beginning of the 20th century.
Some of the archives have recently been digitised.
Earlier analysis of the digitised data was presented by, e.g.,
\citet{kariyappa-pap-96,foukal-96,caccin-et-al-98,worden-et-al-98}.
Usage of these data is until now limited because of the numerous problems
and artifacts affecting the images, such as plate defects and aging,
degradation of instrumentation and observing programmes,
geometrical distortions, photometrical uncertainties etc.
\citep{ermolli-et-al-2007b}.
\citet{ermolli-et-al-2009} have analysed the quality of the data of 3
digitised historic archives from Arcetri, Mt.~Wilson and Kodaikanal
observatories.
They have also been able to fix some of the problems, such as, e.g., disc
eccentricity.
Some others are more difficult to correct.
These include photometric uncertainties (due to, e.g., stray light or
missing photographic calibration) and changes in the image content with time
(due to, e.g., varying wavelength of the observations).

\begin{figure}
   \resizebox{\hsize}{!}{
   \includegraphics{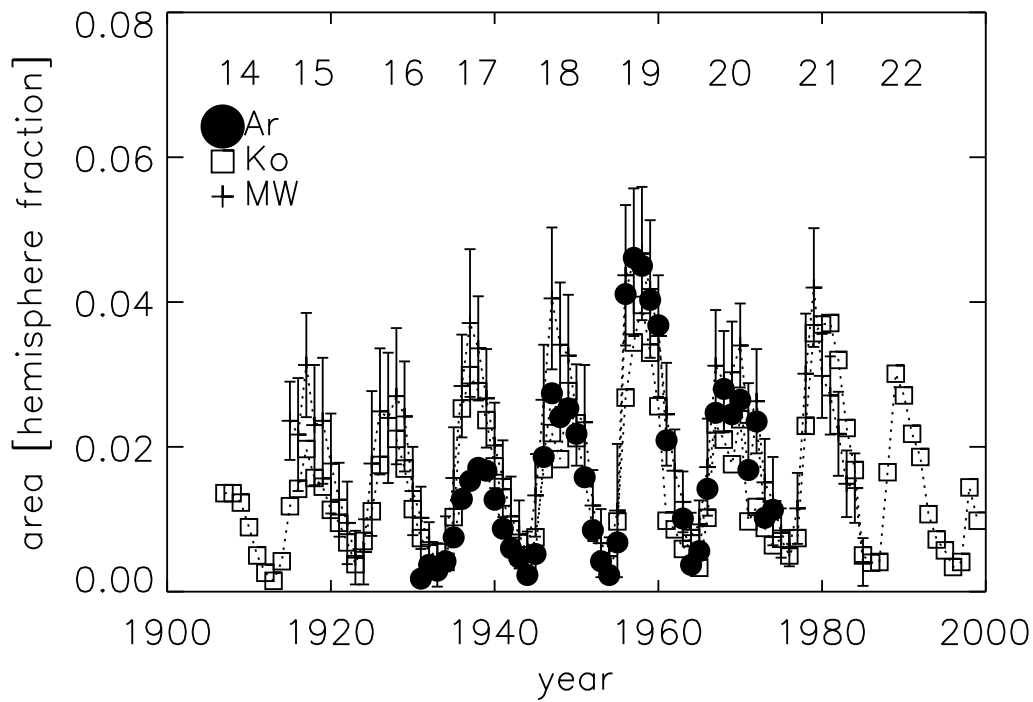}}
   \caption[]{Temporal variation of yearly median values of the plage
coverage measured from the Arcetri, Kodaikanal, and Mt. Wilson series. 
The error bars represent the standard deviation of measured
values over the annual interval; for clarity, they are only shown for the MW
series. Cycle numbers are given at the top of each cycle. From
\citet{ermolli-et-al-2009}
}
   \label{fig:caii}
\end{figure}

\citet{ermolli-et-al-2009} have developed a method of image calibration
independent of the calibration exposures and calculated plage coverage as a
function of time for the 3 analysed archives (Fig.~\ref{fig:caii}).
The Pearson correlation coefficients for each pair of the data sets lie
between 0.85 and 0.93.
The most significant differences are found for cycles 15, 17 and 19, which
are partly explained by the stray-light degradations at the beginning of
each observing programme and by differences in the wavelength of
observations.
Similar results were obtained by \citet{tlatov-et-al-2009}, who also
analysed and compared Kodaikanal and Mt.~Wilson observations (see their
Fig.~4).
These results suggest that caution is needed when using such data  without
careful analysis of their problems and intrinsic instrumental variations.

\section{Solar Activity Over the Holocene}
\label{mill}

Before 1610, no direct proxies of solar magnetic activity are available.
The indirect proxies at hand are the concentrations of the cosmogenic
isotopes $^{10}$Be and $^{14}$C in natural terrestrial archives.
These isotopes are produced when high-energy cosmic rays enter the Earth's
atmosphere and react with nitrogen and other atoms.
The flux of the cosmic rays at the Earth is modulated by the solar open
magnetic field, which makes it possible to reconstruct the open magnetic
flux using the production rates of the cosmogenic radioisotopes.

\citet{solanki-et-al-2004} and \citet{usoskin-et-al-2004,usoskin-et-al-2006a}
have
combined physics-based models for each of the processes connecting $^{14}$C
and $^{10}$Be concentration with sunspot number and thus were able to
reconstructed the sunspot number over the whole Holocene.
This reconstruction has shown that the Sun is currently in a state of
unusually high magnetic activity and that prolonged periods of very low
solar activity similar to the Maunder minimum (grand minima) have regularly
occurred in the past \citep[see also][]{schove-55,siscoe-80,krivsky-84}.
Also periods of higher averaged solar activity, similar to the recent one
(grand maxima) have occurred.
\citet{usoskin-et-al-2007} give a list of all grand minima and maxima since
9500~BC detected in the reconstructed sunspot number record.

The reconstruction of solar irradiance for the Holocene is complicated by the
fact that only a cycle-averaged sunspot number can be derived from the
$^{14}$C and $^{10}$Be concentrations.
Thus the SATIRE-T model described in Sect.~\ref{cent} cannot be applied
directly.
Therefore, first relationships are derived, using the SATIRE-T results,
between the cycle averaged open magnetic flux (the primary solar quantity
obtained from $^{14}$C and $^{10}$Be concentrations) and the total magnetic
flux, the sunspot number and solar irradiance.
These relationships are then used to calculate the solar irradiance over the
Holocene (Vieira et al., in prep.).

\section{Summary}
\label{summary}

As weak as they are (about 0.1\% over the solar cycle), solar irradiance
variations affect the Earth's climate at a level of 0.2$^\circ$K over the
solar cycle \citep{camp-tung-2007}.
The role of the total solar irradiance might be significantly reinforced by
the wavelength dependence of irradiance variations.
The amplitude of the variations is 1 to 3 orders of magnitude higher in the
UV part of the spectrum than in the visible or IR.
At the same time, solar UV radiation is an important driver of chemical and
physical processes in the Earth's upper atmosphere.

Reliable assessment of solar forcing on the Earth's climate is still
plagued, among other factors, by a shortage of reliable and sufficiently long
irradiance records.
The time series of direct space-borne measurements of solar irradiance is
only 3 decades long (in case of spectral irradiance, even shorter).
Therefore we have recently focussed our
main attention on tests of the reconstructed irradiance
against the newest spectral data, improvement of the models in the UV and
appraisal of the magnitude of the secular trend in irradiance.

We have reconstructed
solar spectral irradiance at 115--160000~nm  back to 1947.
Reconstruction of solar total irradiance goes back to 1610 and suggests a
value of about 1--1.5~Wm$^{-2}$  for an increase in the cycle-averaged TSI
between the end of the Maunder minimum and the last 50~years, which is
significantly lower than previously accepted but agrees with other recent
estimates.
First steps have also been made towards reconstructions of solar total and
spectral irradiance on time scales of millennia.

\bigskip
{\bf Acknowledgments.}
This work was supported by the Deutsche For\-schungs\-ge\-mein\-schaft, DFG project
number SO~711/2 and by the WCU grant No.~R31-10016 funded by the Korean
Ministry of Education, Science and Technology.
We thank the International Space Science Institute (Bern) for
hosting meetings of the
international team on `Interpretation and modelling of SSI measurements'.

\end{document}